\documentstyle[11pt,aasms,tighten,flushrt]{article}
\def \cm{~\rm{cm}}
\def \s{~\rm{s}}
\def \km{~\rm{km}}

\def \g{~\rm{g}}
\def \AU{~\rm{AU}}

\def \G{~\rm{G}}
\def \lesssim{\mathrel{<\kern-1.0em\lower0.9ex\hbox{$\sim$}}}
\def \gtrsim{\mathrel{>\kern-1.0em\lower0.9ex\hbox{$\sim$}}}

\begin{document}


\title{
EXTRASOLAR PLANETS AND THE ROTATION AND AXISYMMETRICAL
MASS LOSS OF EVOLVED STARS }

\author{ 
Noam Soker
\affil{
Department of Physics, University of Haifa at Oranim\\
Oranim, Tivon 36006, ISRAEL \\
soker@physics.technion.ac.il }}

\bigskip


\centerline {\bf ABSTRACT}

 I examine the implications of the recently found extrasolar planets
on the planet-induced axisymmetrical mass loss model for the
formation of elliptical planetary nebulae (PNs).
 This model, which was developed in several earlier papers by
the author and a few collaborators, attributes low departure from
spherical mass loss of upper asymptotic giant branch (AGB) stars
to envelope rotation which results from deposition of
planet's orbital angular momentum.
 Since about half of all planetary nebulae are elliptical, i.e.,
have low equatorial to polar density contrast,
it was predicted that $\sim 50 \%$ of all sun-like stars
have Jupiter-like planets around them, i.e., a mass about equal
to that of Jupiter, $M_J$, or more massive.
 In light of the new finding that only $\sim 5 \%$ of sun-like stars
do have such planets, and a newly proposed mechanism for
axisymmetrical mass loss, the cool magnetic spots model, I
revise this prediction.
 I predict that indeed $\sim 50 \%$ of PNs progenitors do have
close planets around them, but the planets can have much lower
masses, as low as $\sim 0.01 M_J$, in order to substantially
spin-up the envelopes of AGB stars.
 To support this claim I follow the angular momentum evolution of
single stars with main sequence mass in the range of $1.3-2.4 M_\odot$,
as they evolve to the post-AGB phase.
 I find that single stars  rotate much too slowly to possess any
significant non-spherical mass loss as they reach the upper AGB.
 It seems, therefore, that planets, in some cases even Earth-like planets,
are required to spin-up the envelope of these AGB stars for them to
form elliptical PNs.
 The prediction that on average several such planets orbit each star,
as in the solar system, still holds. 




\section{INTRODUCTION}

 About forty sun-like stars (i.e., main sequence mass of
$M_{\rm ms} \lesssim 1.3 M_\odot$) are presently known to have
close planets around them (for recent studies see, e.g.,
Marcy \& Butler 2000; Santos {\it et al.} 2000).
  In several billions years these stars will start their evolution along
the red giant branch (RGB), reaching radii of $\sim 100 R_\odot$
on the upper RGB.
 Most of the known extrasolar planets will be engulfed by
their evolving parent (central) star.
 Those planets which survive the RGB evolution will be engulfed
by their central stars as the latter evolve to become asymptotic giant
branch (AGB) stars, when they reach radii of $\gtrsim 200 \AU$.
 Even if the initial separation of a planet is larger than the maximum
radius of the star on the AGB, tidal forces will cause all known
extrasolar planets to spiral-in into the envelope of their evolved
parent stars (see eq. 6 by Soker 1996).
 The evolution of planets inside the envelopes of evolved stars
was studied long before the detection of extrasolar planets
(e.g., Eggleton 1978; Livio 1982; Livio \& Soker 1984;
Harpaz \& Soker 1994) as well as after several extrasolar planets
have been detected (e.g., Siess \& Livio 1999a,b).
 The most obvious effect of planets will be the
spinning-up of the evolved stellar envelopes.
 This is because RGB and AGB single stars are expected to rotate
very slowly, so that even an Earth-like planet may more than double
the rotational velocity of AGB stars (see $\S 3$ below).
 In several earlier papers (e.g., Soker \& Harpaz 1992; Soker 1996, 1997,
and references therein),
I suggested that the spinning-up of evolved stars by substellar objects
(i.e., brown dwarfs and planets) may lead to axisymmetrical, rather than
spherical, mass loss on the upper AGB.
 As a result of this axisymmetrical mass loss, the descendent
planetary nebulae (PNs) will be moderately elliptical rather than spherical.
By moderately elliptical PNs I refer to those with a small to moderate
deviation from sphericity, and not to PNs which contain lobes (i.e.,
bipolar PNs), rings (i.e., extreme elliptical PNs), and other
structures with large departure from sphericity, and which I believe
require stellar companions (Soker 1997).   

The planet-induced axisymmetrical mass loss model for the formation
of elliptical PNs had led to three major predictions:
(1) Many planets will be more massive and closer to their parent stars
than Jupiter is (Soker 1994, 1996).
(2) For many stars to engulf a planet at their late evolutionary stages
with a high probability, several substellar objects must
be present in most of the systems (Soker 1996).
(3) About half of all progenitors of PNs have planetary systems,
containing Jupiter-like, or more massive, planets (Soker 1997).
This prediction is based on the notion that singly evolved stars
rotate too slowly when they become AGB stars, and hence have
spherical mass loss.
Since most PNs are axisymmetrical rather than spherical, but only
$\sim 30-50 \%$ have stellar companions in the appropriate
range of masses and orbital separations, most other PNs' progenitors
were spun-up by planets (Soker 1997).
 
 The first two predictions were made before extrasolar planets were
found, and the third was made when only a small number
of extrasolar planets were known.
 Presently, more than 40 extrasolar planets around solar-like stars
are known.
 These planets show indeed that planets more massive than Jupiter
and at much closer orbits do exist.
The first prediction was therefore confirmed.
The second prediction has not been disproved or confirmed yet,
since the detection sensitivity is too low for any
meaningful conclusion.
 Only one system, upsilon Andromedae, was found to be composed
of at least three planets (Butler {\it et al.} 1999).
 The third prediction was disproved.
 With current detection limits, only $\sim 5 \%$ of the stars in the
different samples were found to have Jupiter-like planets
(Marcy \& Butler 2000), and a very small number of systems
do have brown dwarfs (Halbwachs {\it et al.} 2000).
 Even if I consider the detection sensitivity and allow for orbital
separations twice as large as the detection limit and masses down to
$0.3 M_J$, where $M_J$ is the mass of Jupiter, the fraction
of such extrasolar planetary systems  is $ \lesssim 10 \%$ among
all sun-like stars.

 The solution to the conflict between the third prediction above and
the properties of known extrasolar planets may be one of the following.
(a) If most stars of masses above those for which planets have been
searched, i.e., $M_{\rm ms} \gtrsim 1.3 M_\odot$, possess
Jupiter-like planets, and if these more massive stars form most PNs,
while only a small fraction of stars having main sequence mass of
$M_{\rm ms} \lesssim 1.3 M_\odot$ form PNs,
then the statistics of known extrasolar planets has only slight
implication for the progenitors of PNs.
  Allen, Carigi, \& Peimbert (1998) argue that most stars with
$M_{\rm ms} \lesssim 1.3 M_\odot$ do not form PNs.
They claim that only $\sim 50 \% $ of stars with $M_{\rm ms} = 1.3 M_\odot$
form PNs, with decreasing probability for lower masses.
 One problem with this solution is that many of the stars
in the known extrasolar planetary systems will {\it not} form PNs at all.
 Instead, I expect them to lose most of their envelope on the RGB,
becoming blue horizontal branch (BHB) stars, and then fading as WD without
an observable nebula.
The rest of the sun-like stars, those that have no Jupiter-like
planets, are more likely to form PNs.  
This solution seems unlikely, therefore.
\newline
(b) If singly evolved AGB star rotate fast enough to induce
axisymmetrical mass loss.
 Following the results of $\S 2$ below, where I follow the evolution of
the rotation velocity of AGB stars, I consider this possibility
to be very unlikely (section 3).
\newline
(c) Planets of masses as low as a few times Earth mass, i.e.,
$M_p \sim 0.01 M_J$, are sufficient to spin-up AGB stars for the
planet-induced axisymmetrical mass loss model to work.
In $\S 3$ I suggest that this is indeed the case,
and that $\sim 50 \%$ of PNs progenitors do have planetary systems,
but in most cases the most massive planet has a mass in the range of
$0.01 M_J \lesssim M_p \lesssim 0.1 M_J$.
 Summary of main results and predictions are in $\S 4$.
 
\section{ANGULAR MOMENTUM EVOLUTION}

 I concentrate on stars with main sequence mass in the range of
$1.3 M_\odot < M_{\rm ms} < 2.4 M_\odot$.
 The major reason is that in this mass range the transition from slow
main sequence rotators to fast rotators occurs (e.g., Wolff \& Simon 1997).
 These stars will clearly demonstrate the evolution of angular momentum,
while avoiding some uncertainties with lower mass stars.
 The uncertainties with lower mass stars are in the total mass that
is lost during the RGB, and later during the early AGB.
As is evident from BHB stars in globular clusters, low mass
stars may lose almost their entire envelope on the RGB
(Dorman, Rood, \& O'Connell 1993;  D'Cruz {\it et al.} 1996).
   Stars having $M_{\rm ms} \sim 1-1.3 M_{\rm ms}$ will
retain most, but still lose a substantial fraction, of their envelopes
on the RGB as well.
 Following in detail only the $1.3-2.4 M_\odot$ mass range is sufficient
for the present goals since it seems that a large fraction of all
PNs result from these stars (Allen {\it et al.} 1998).
 To make possible analytical integration of the angular momentum loss,
I approximate the observations presented by Wolf \& Simon (1997;
e.g., their figures 3 and 4) for the average angular velocity
of these main sequence stars by
\begin{eqnarray}
v = 200 (M_{\rm ms} -1.2) \km \s^{-1},
\end{eqnarray}
where the mass is given in solar mass units.
 This approximation takes into account the inclination effect ($\sin i$),
and the possibility that the massive stars in this range retain some extra
angular momentum in their cores.
 Therefore, this velocity function is somewhat higher than $v \sin i$
given in figure 4 of Wolff \& Simon (1997) in most of the mass range. 
 To obtain a simple expression for the total angular momentum, I take the
radii of these stars to be $R_{\rm ms}=M_{\rm ms}^{0.75}$, where masses
and radii are in solar units.
 I also take the ratio of the moment of inertia to $M_{\rm ms} R_{\rm ms}^2$
to be $\sim 0.07$, as for the sun.
 Using these two approximations I find for the average total angular
momentum of these stars on the main sequence 
\begin{eqnarray}
J_{\rm ms} \simeq M_{\rm ms}^{1.75}  (M_{\rm ms} -1.2) J_J, \qquad
{\rm for} \qquad 1.3 < M_{\rm ms} < 2.4,
\end{eqnarray}
 where the masses are in solar mass units, and
$J_J=1.9 \times 10^{50} \g \cm^2 \s^{-1}$ is the orbital
angular momentum of Jupiter.

 For the angular momentum evolution I follow an earlier paper
(Soker \& Harpaz 2000), and use the same notations and derivation of
the angular momentum loss. 
  Not considering magnetic influence beyond the stellar surface,
and assuming a solid body rotation through the stellar envelope,
the angular momentum loss rate from stars is
\begin{eqnarray}
\dot J_{\rm wind} = \beta \omega R^2 \dot M,
\end{eqnarray}
where $\omega, J, R, M$ are the stellar angular velocity,
angular momentum, radius, and mass, respectively,
and $\beta$ depends on the mass loss geometry.
 For a constant mass loss rate per unit area on the surface $\beta =2/3$,
while for an equatorial mass loss $\beta=1$. 
 The angular momentum of the star is $J_{\rm env}=I \omega$, where $I$
is the moment of inertia given by
\begin{eqnarray}
I = \alpha M_{\rm env} R^2,
\end{eqnarray}
where $M_{\rm env}$ is the envelope mass, and 
I neglect the core's moment of inertia relative to that of the envelop
and the change in the core mass at late AGB stages.
 Dividing equation (3) by equation (4) multiplied by $\omega$, gives 
\begin{eqnarray}
\frac {d \ln J_{\rm env}}{d \ln M_{\rm env}} = 
\frac {\beta}{\alpha (M_{\rm env})} \equiv \delta .
\end{eqnarray}

 In order to integrate equation (5) along the evolution of the star,
we turn to find the variation of $\alpha$ with the envelope mass.
 We only consider mass loss on the upper RGB and upper AGB,
where most of the mass loss occurs.
  Soker \& Harpaz (2000) find the value of $\alpha$ on the RGB
for sun-like stars to be (if the envelope mass is not too low)
$\alpha_{\rm RGB} \simeq 0.1$.
 For spherical mass loss $\beta = 2/3$, hence on the RGB
I take $\delta_{\rm RGB} = 6.7$.
 We expect the envelope mass to be lost on the RGB to be $\sim 0.2 M_\odot$.
Along the AGB we find from the models presented by Soker \& Harpaz (1999;
their figures 1-5; note that the density scale in figures 1-5 is
lower by a factor of 10; the correct density scale is in their fig. 6)
\begin{eqnarray}
\alpha^{-1} \simeq 10 M_{\rm env}^{0.3} \qquad {\rm for}
\qquad 0.01 \lesssim M_{\rm env} < 0.5 \quad {\rm on~ upper~ AGB},
\end{eqnarray}
again, all masses are in solar mass units.
 At earlier AGB stages when $M_{\rm env} >0.5 M_\odot$ I take 
$\alpha = 8.25$.
 For the core mass I take $M_c = 0.6 M_\odot$ for all the calculations,
since most of the mass loss occurs on the upper AGB, during a phase
when the core mass does not increase substantially.
 Since the core mass on the RGB is smaller than $0.6 M_\odot$, using
a core mass of $0.6 M_\odot$ means a lower envelope mass, hence for a
given mass loss the angular momentum loss will be overestimated.
 To compensate, I take a lower value of $\delta$, namely
$\delta = 5.5$ instead of $6.7$.
 Although crude, this approximation simplifies the calculation
 substantially, while still being adequate for the present goals.
 As noted earlier, for lower mass stars, $M_{\rm ms} <1.3 M_\odot$,
these approximations are not applicable, since the mass loss on the
RGB is significant. 

 Using these values of $\alpha$ and taking spherical mass loss $\beta=2/3$,
I derive the following approximation for $\delta$ (defined in equation 5)
\begin{eqnarray}
\delta \simeq 5.5 (M_{\rm env}/0.5)^{0.3} \qquad {\rm for}
\qquad 0.01 \lesssim M_{\rm env} < 0.5 \\ \nonumber
 5.5 \qquad {\rm for} \qquad M_{\rm env} \geq 0.5 
\end{eqnarray}
 Substituting $\delta$ from last equation allows analytical
integration of equation 5. 
 For envelope masses of $M_{\rm env}<0.5 M_\odot$ (again, 
for an assumed core mass of $0.6 M_\odot$ during the stage
when most of the mass is being lost), the angular
momentum is 
\begin{eqnarray}
{ {J}\over{J_J}} \simeq M_{\rm ms}^{1.75}  (M_{\rm ms} -1.2) 
\left( {M_{\rm ms}-0.6}\over{0.5}\right)^{-5.5}
\exp[-22.6(0.5^{0.3}-M_{\rm env}^{0.3})] 
\qquad 0.01 \lesssim  M_{\rm env} < 0.5. 
\end{eqnarray}
 The angular momentum as a function of main sequence mass is plotted
 on Figure 1 for envelope masses of
$M_{\rm env} =0.4$, $0.2$ and $0.1 M_\odot$.
 The  peak near $M_{\rm ms}=1.4 M_\odot$ is not real. 
It seems rather that the angular momentum on the upper AGB 
does not depend much on the initial mass for stars 
with $M_{\rm ms} \lesssim 1.4 M_\odot$. 
  Even for the sun we get similar values. 
 Since some of the early assumptions do not hold for a star with 
$M_{\rm ms} \lesssim 1.3 M_\odot$, I assume the following instead.
 A sun-like star loses $0.2 M_\odot$ on the RGB, and then another
$0.05  M_\odot$ on the early AGB. 
 With an average core mass of only $0.4 M_\odot$ on the RGB, and 
$\delta=6.7$ as mentioned above, the sun will retain only 
$(0.4/0.6)^\delta = 0.066$
of its angular momentum when leaving the RGB. 
 Then with $M_{\rm core} =0.6 M_\odot$ and $\delta=5.5$ on the early AGB,
the angular momentum will be reduced by another factor of $5$
when the envelope reaches mass of $M_{\rm env}=0.15 M_\odot$.
 At this stage the sun's angular momentum will be $\sim 0.014$ times
it main sequence value. Since the angular momentum of the sun is
$0.01 J_J$, the sun will have $J \simeq 10^{-4} J_J$ when 
$M_{\rm env}=0.15 M_\odot$.
This is very similar to the value for stars with $M_{\rm ms}=1.4 M_\odot$.
 Therefore, it seems that the lines in Figure 1 will become horizontal 
when continued to the left down to $M_{\rm ms} \simeq 1 M_\odot$. 

The angular velocity is given by $J=\omega I$.
Taking $I$ from equation (4) and $\alpha$ from equation (6)
gives
\begin{eqnarray}
{{\omega}\over{\omega_{\rm Kep}}}
= 1.1 \times 10^{-4}
M_\ast^{-1/2}
\left( {{M_{\rm env}}\over{0.1}} \right)^{-0.7}
\left( {{R}\over{\AU}} \right)^{-1/2}
\left( {J}\over{10^{-3} J_J} \right)
\qquad 0.01 \lesssim M_{\rm env} < 0.5, 
\end{eqnarray}
where masses are in solar units,
$\omega_{\rm Kep}=(G M_\ast/R^3)^{1/2}$ is the Keplerian angular
velocity on the stellar equator $R$, and $M_\ast=0.6+M_{\rm env}$
is the total stellar mass, where $M_{\rm env}$ is in solar units.
 This equation is correct for any envelope angular momentum, and
is not restricted to a single star evolution as is equation (8).

 Figure 2 shows the evolution of the angular velocity (solid line),
 in units of $\omega_{\rm Kep}$, and
the angular momentum (dashed line), in units of $J_J$,
as a function of the envelope mass along the upper AGB for a single star
with $M_{\rm ms}=1.8 M_\odot$.
 The angular momentum is according to equation (8), while the angular
velocity is according to equation (9), with $J$ from equation
(8), and for $R=1 \AU$, hence the lower limit on the envelope mass
of $M_{\rm env} = 0.03 M_\odot$, below which the stellar radius
decreases much below $1\AU$. 
 We note the fast decrease of the angular velocity as envelope
mass decreases due to mass loss.

\section{THE CONSTRAINTS ON ANGULAR VELOCITY AND PLANETS}

 For the formation of elliptical PNs the question is what
angular velocity should an AGB envelope have in order to possess
some mass loss rate contrast between equatorial and polar directions.
 For mechanisms which are based on dynamical effects of rotation,
i.e., due to the centrifugal force,
the envelope should rotate at $\omega \gtrsim 0.1 \omega_{\rm Kep}$
(e.g., Dorfi \& H\"ofner 1996; Garcia-Segura {\it et al.} 1999).
 These models require, therefore, that the AGB star be spun-up by
stellar companions (Soker \& Harpaz 1999),
and are not relevant to the present discussion.
 This holds true for mechanisms based on dynamical effects of 
magnetic fields (e.g., Chevalier \& Luo 1994; Garcia-Segura 1997;
Pascoli 1997).
 Noting that the envelope of a singly evolved AGB star rotates very
slowly, some works examine the role of a fast rotating core.
 Garcia-Segura {\it et al.} (1999) assume that the core retains a high
rotational velocity, and only at the very end of the AGB does the
core transfer its angular momentum to the envelope; by this stage
the envelope contains a very low mass, and hence is efficiently
spun-up by its coupling to the core.
 This scenario was criticized in a previous paper
(Soker \& Harpaz 1999; their $\S 3.2$).
  Pascoli (1997) proposes that the asymmetric mass loss is caused by
a strong magnetic field which is amplified in the core,
and then transported to the AGB stellar surface.
 Therefore, only the core and the envelope close to the core
are required to rotate fast, so that the total stellar angular
momentum is very small.
 There are several problems I see with Pascoli's (1997) model.
 Among them are the too strong magnetic pressure on the stellar surface,
which is about equal to the thermal and convective pressure there.
 Others problems are the too strong magnetic field and too fast rotation
assumed on the core's surface, $10^6 \G$ and $\omega_c=10^{-2} \s^{-1}$
(i.e., orbital period of $\sim 10$ minutes).
 Single white dwarfs have much weaker magnetic fields;
only $\sim 4 \%$ of all white dwarfs have a magnetic field
of strength $B \gtrsim 3 \times 10^4 \G$,
and they are much slower rotators, having orbital periods of $\gg 1 hr$
(Schmidt \& Smith 1995).
  Despite the problems with the scenario proposed by Pascoli (1997),
the idea that the axisymmetric magnetic field on the AGB surface results
from an amplification close to the core deserves further study, in
particular in models which require much weaker surface magnetic fields.
 Such is the cool magnetic spots model (Soker \& Harpaz 1999; Soker 2000),
where it is assumed that a weak magnetic field forms cool stellar spots,
which facilitate the formation of dust closer to the stellar surface,
hence increasing the mass loss rate.
 If spots due to the dynamo activity are formed mainly near the equator,
an enhanced equatorial mass loss is obtained.
 One problem still remains with the idea of magnetic field transport
from the core, since AGB stars have strong convection,
which extends from the surface down to $\sim 1 R_\odot$.
 It is not clear that such a deep convective envelope with strong
convection can maintain the axisymmetric structure of the
core's magnetic field without any envelope rotation.
 It is more likely that the convection will smear the magnetic field,
so that on average the surface magnetic field will be spherical.
 
 It seems therefore that some envelope rotation is required.
 The role of the slow envelope rotation is only to trigger another
process which directly causes the axisymmetrical mass loss.
 In the cool magnetic spots model the role of the rotation is mainly to
shape the magnetic field into an axisymmetrical configuration
(on average), and it may operate efficiently even for an envelope
rotating as slowly as $\omega \sim 10^{-4} \omega_{\rm Kep}$
(Soker \& Harpaz 1999).
 For such slow envelope rotations, very low mass planets,
down to $\sim 0.01 M_J$, are sufficient, if they enter the
AGB envelope at late stages.
 For example, a planet of mass $0.01 M_J$ at an orbital separation
of $2 \AU$ has an angular momentum about equal to
that of an AGB star with envelope mass of $M_{\rm env} =0.4 M_\odot$
which had a main sequence mass of $M_{\rm ms} =1.8 M_\odot$.
 If such a planet enters the envelope when $M_{\rm env}=0.2 M_\odot$,
for example, it will increase the AGB envelope angular momentum
by a factor of $\sim 30$.  
 Taking the solar evolution to the AGB as described in section 2,
I find that when the envelope mass becomes $0.15 M_\odot$,
the angular momentum of the AGB sun is $\sim 10^{-4} J_J$,
or $\sim 0.1$ the angular momentum of Earth.
 By that time the orbital separation will be $1.33 \AU$
(or $290 R_\odot$).
  If the sun at this stage goes through a helium shell flash, so that
the radius increases, say, to $\sim 1.3 \AU$
(e.g., Boothroyd \& Sackmann 1988; note that they use too short a mixing 
length, and consequently their radii are overestimated by a factor of 
$\sim 2$ on the RGB and AGB), then another $\sim 10 \%$ increase
during the maximum radius in the pulsation cycles may reach the location of
Earth, causing the Earth to spiral inside the solar envelope.
 A detailed analysis of the evolution of the Earth-sun system, until the
sun leaves the AGB,  for different assumptions and models,
is given by Rybicki \& Denis (2000).
  As a result of the deposition of the Earth's orbital momentum,
the solar envelope will rotate $\sim 10$ times faster,
or at $\omega \simeq 10^{-4} \omega_{\rm Kep}$.
 If this occurs indeed in about 7 billion years, then the Earth may be
responsible for the PN of the sun being elliptical rather
than spherical.
 However, it is not clear that the sun will engulf the Earth,
or that it will form a PN at all (Rybicki \& Denis 2000).

 To have a high probability that a planet will enter the AGB
envelope at late stages, i.e., for it to occur in many stars,
two things should happen.
First on average there should be several planets around each
star (as is the case in the solar system), and second,
there should be a fast and significant increase of the stellar radius
on the upper AGB.
 Numerical simulations of AGB stars show that after thermal pulses
(helium shell flashes) on the upper AGB, the envelope increases
by $\sim 20-30 \%$ (e.g.,  Boothroyd \& Sackmann 1988).
 This is in addition to the increase in the average AGB stellar radius
as the core mass increases.
 So the second condition is fulfilled for upper AGB stars.
 The first condition is a requirement, and hence a {\it prediction},
of the planet-induced axisymmetrical mass loss model for the formation
of elliptical PNs.
 The new addition of the present paper is the relaxation of the
minimum mass demand on planets from $\sim 1 M_J$
(Soker 1996) to $\sim 0.01 M_J$.
The motivations for reducing the lower mass limit are
the new finding that only $\sim 5 \%$ of sun-like stars have
Jupiter-like planets around them, and a new model for
axisymmetrical mass loss, the cool magnetic spots model, which
was constructed to work for very slowly rotating AGB stars by the
author and a few collaborators (e.g., Soker 2000).

 Finally, it should be noted that many of the known sun-like
stars that have planets around them will not form PNs at all.
This is since their orbiting planet will spin-up the envelope
and deposit energy already on the stellar RGB, hence mass loss
on the RGB is expected to be high, and most of the stellar envelope
will be lost already on the RGB.
 No observable nebula will be formed.
 Hence, while in most cases planet companions will lead to the
formation of an elliptical rather than a spherical PN, in some cases
Jupiter-like planets in close orbits around low mass stars,
$M_{\rm ms} \lesssim 1.2 M_\odot$, will prevent the stars
from forming a PN.
 This process, of planet-induced envelope loss on the RGB, is interesting
for globular clusters where there are many blue horizontal branch (BHB)
stars.
 The BHB stars are thought to result from stars which lost most of
their envelope on the RGB (e.g., Dorman {\it et al.} 1993;
D'Cruz {\it et al.} 1996).
 It is not clear yet what causes this higher mass loss rate on the
RGB, but one possibility is the presence of close planets
(Soker \& Harpaz 2000).
 The question here, again, is what is the minimum planet's mass
required to substantially enhance the mass loss rate.
 Some hints come from the angular momentum of BHB stars, which
indicate that planets with masses of $\sim 0.1 -10 M_J$ are
sufficient to induce the enhanced mass loss rate (Soker \& Harpaz 2000).
 The planet-induced high mass loss rate model predicts that main sequence
stars in globular clusters with no BHB will have no massive planets
around them, while many main sequence stars in globular clusters with
many BHB will have planets around them.
 It is interesting that in a recent work Brown {\it et al.} (2000) report
that no planets were found around main sequence stars in the globular
cluster 47 Tuc (NGC 104).
 This globular cluster contains no (or only a few) BHB (Rich {\it et al.}
1997), and therefore I do not expect the stars in this globular cluster
to have massive and close planets around them.
It will be interesting to repeat such observations for globular clusters
which have many BHB, for which I do expect the presence of planets
around many main sequence stars.
    
\section{SUMMARY}
         
 The goal of the present work was to update some earlier results
of the author and collaborators regarding the planet-induced
axisymmetrical mass loss model for the formation of elliptical PNs.
 To do so I followed the angular momentum evolution of single stars
in the main sequence mass range of $1.3 \leq M_{\rm ms} \leq 2.4 M_\odot$
(section 2).
 The main results of the present paper can be summarized as follows. 
\newline
 (1) Single stars will rotate extremely slowly when they
reach the upper AGB (Figs. 1 and 2).
Therefore, they are not likely to form elliptical PNs, but
spherical PNs.
\newline
(2) If a planet with a mass of $M \gtrsim 0.01 M_J$ is engulfed by the
star as it reaches the AGB, the star will be spun-up substantially
by the deposition of the planet's angular momentum.
 The rotation by itself will not deform the AGB wind, but may trigger
another process that will lead to axisymmetrical mass loss,
e.g., weak magnetic activity, as in the magnetic cool spots model
(Soker 2000).
 The required angular velocity in that model is only $\sim 10^{-4}$
times the Keplerian velocity on the stellar surface.
\newline
(3) Stars which have close stellar companions are likely to form bipolar
PNs, i.e., those with two lobes and an equatorial waist between the
lobes, or elliptical PNs with extreme equatorial to polar
density contrast, e.g., a ring-like PN.
 However, these systems can add up to no more than $\sim 50 \%$ of
all PNs, while $\sim 90 \%$ of all PNs are aspherical.
 This led me in earlier papers (e.g., Soker 1997) to suggest that
$\sim 50 \%$ of all progenitors of PNs have Jupiter-like planets
around them.
 This is in contradiction with recent findings that only
 $\sim 5 \%$ of sun-like stars have Jupiter-like planets around
them at close orbits (Marcy \& Butler 2000).
 I still maintain my claim that $\sim 50 \%$ of all progenitors of
PNs have planets around them, but now I suggest that the lower
mass on the planets' masses be reduced to $\sim 0.01 M_J$.
\newline
(4) In order for such low mass planets to substantially spin-up
the stellar envelope, they should enter the envelope when the
star reaches the upper AGB.
 This ``fine-tuning'' can be avoided if there are several planets
on average around each star, as is the case in the solar system,
so that one of them is engulfed when the star reaches the upper AGB.
 Therefore I retain earlier predictions (Soker 1996) that on
average several planets are present around $\sim 50 \%$ of
progenitors of PNs.
\newline
(5) I argue that most known sun-like stars that have planets around them
will not form PNs at all, but the deposition of planets' orbital
angular momentum and energy will cause most, or even all, of the envelopes
of these stars to be lost already on the RGB (Soker \& Harpaz 2000).
 This is the case for main sequence stars with
$M_{\rm ms} \lesssim 1.2 M_\odot$ with a close Jupiter-like planet
around them, as are most of the presently known extrasolar planets.
 These stars will not reach the upper AGB after the horizontal branch,
and no observable nebulae will be formed.
 This scenario should be examined by searching the main sequence stars
of globular clusters with many blue horizontal branch stars 
for close planet companions.

\acknowledgments
This research was supported in part by grants from the 
Israel Science Foundation and the US-Israel Binational Science Foundation.

\bigskip

{\bf FIGURE CAPTIONS}

\noindent {\bf Figure 1:}
 The angular momentum on the upper AGB as a function of main
sequence mass, for envelope masses of
$M_{\rm env} =0.4$, $0.2$ and $0.1 M_\odot$ as indicated.
$J_J$ is the orbital angular momentum of Jupiter, and
envelope masses are in units of $M_\odot$.
 The  peak near $M_{\rm ms}=1.4 M_\odot$ is not real, but 
the angular momentum on the upper AGB 
does not depend much on the initial mass for stars 
with $M_{\rm ms} \lesssim 1.4 M_\odot$. 

\noindent {\bf Figure 2:}
 Evolution of the angular momentum (eq. 8) and angular velocity
(eq. 9 with $R=1 \AU$) as a function of the mass left in the
AGB stellar envelope.
 Angular momentum is in units of Jupiter's orbital angular
momentum, and angular velocity in units of the Keplerian angular
velocity on the stellar equator.
The assumption of $R=1 \AU$ is not accurate for post-AGB stars,
i.e., very low envelope mass, hence the plots do not continue
below $M_{\rm env} =0.03 M_\odot$.

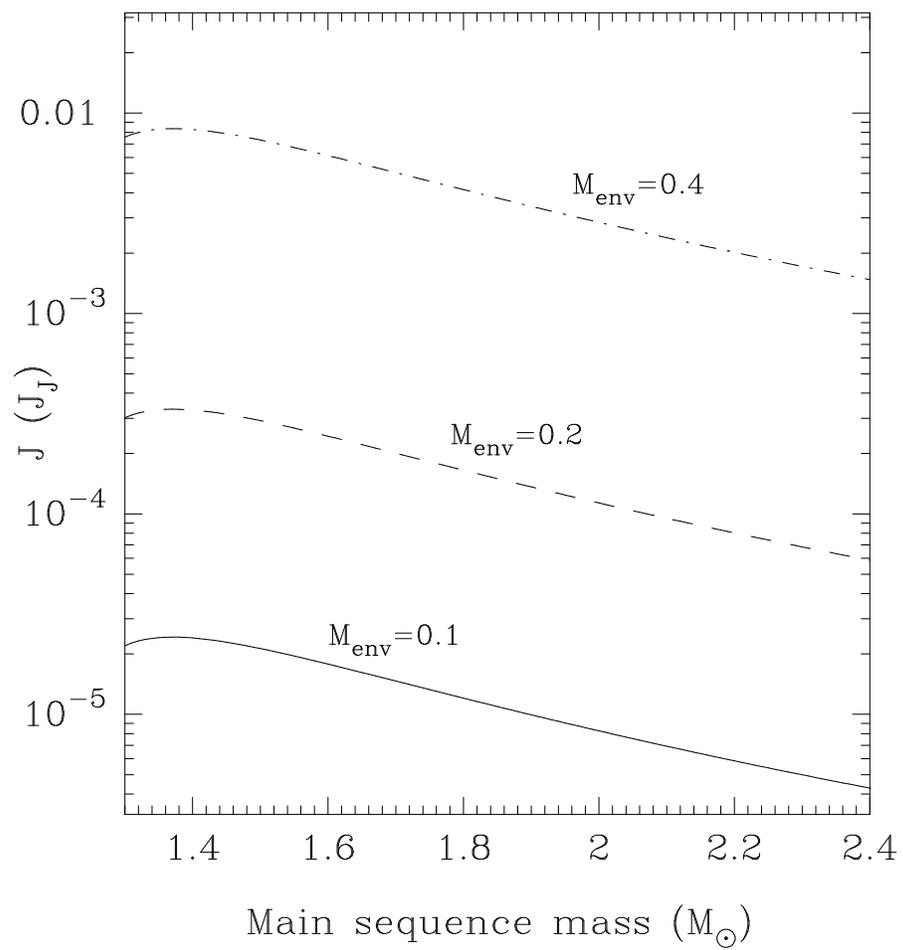
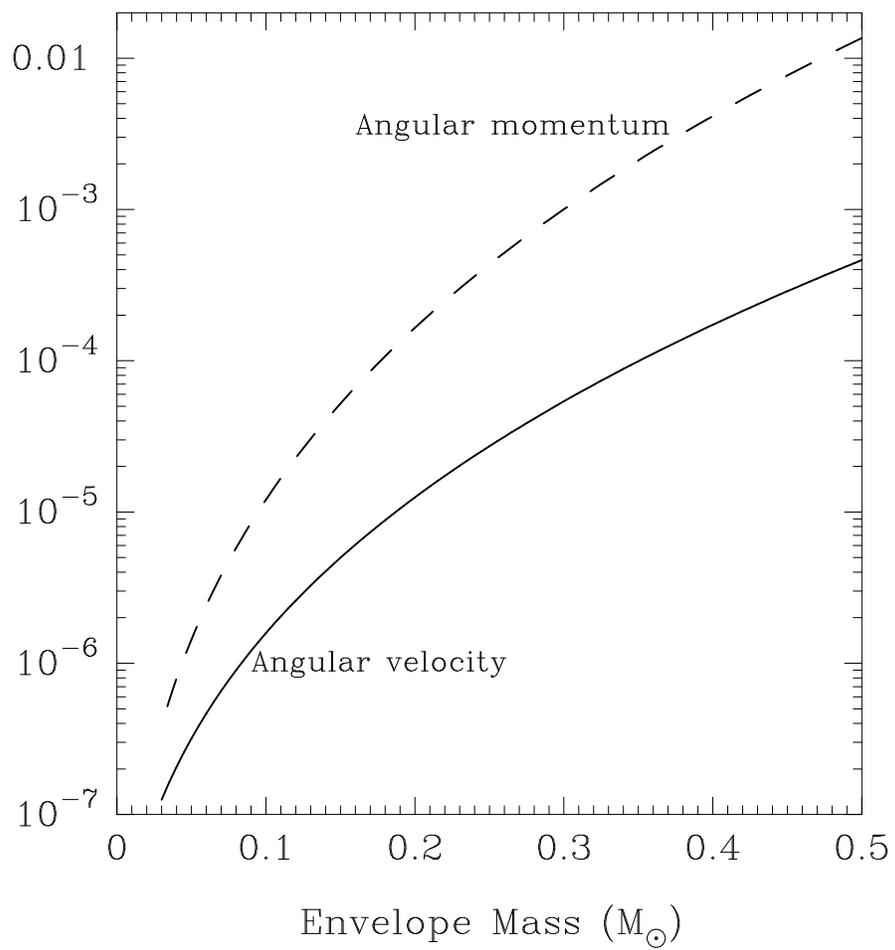
\end{document}